\documentclass[english]{IEEEtran}
\usepackage[T1]{fontenc}
\usepackage[latin9]{inputenc}
\usepackage{color}
\usepackage{verbatim}
\usepackage{amsmath}
\usepackage{amssymb}
\usepackage{wasysym}

\makeatletter
\@ifundefined{date}{}{\date{}}
\usepackage{ifthen}

\newtheorem{theorem}{Theorem}
\newtheorem{remark}{Remark}

\newtheorem{algorithm}{Algorithm}

\newtheorem{example}{Example}

\newcommand{\vs}{\vspace}
\newcommand{\sis}{\setlength{\itemsep}}

\def\acc{0}

\makeatother

\usepackage{babel}
\begin{document}
\title{Quantum optimization for\\Nonlinear Model Predictive Control}
\author{Carlo Novara, Mattia Boggio, Deborah Volpe\thanks{The Authors are with Politecnico di Torino, Dept. of Electronics and
Telecommunications, Italy. Email: \{carlo.novara, mattia.boggio, deborah.volpe\}@polito.it}}

\maketitle
\textbf{Abstract.} %
Nonlinear Model Predictive Control (NMPC) is a general and flexible
control approach, used in many industrial contexts, and is based on
the online solution of a nonlinear optimization problem. This operation
requires in general a high computational cost, which may compromise
the NMPC implementation in ``fast'' applications, especially if a
large number variables is involved. To overcome this issue, we propose
a quantum computing approach for the solution of the NMPC optimization
problem. Assuming the availability of an efficient quantum computer,
the approach has the potential to considerably decrease the computational
time and/or enhance the solution quality compared to classical algorithms.

\section{Introduction}

\ifthenelse{\acc=1}{}{

\subsection{Motivations and contributions}

}

Nonlinear Model Predictive Control (NMPC) is a widely-used control
approach for complex nonlinear systems, see \cite{MaRaRaSc00,FiAl02,Camacho07,Rawlings2019}
and the references therein. The approach provides optimal solutions
(over a finite time interval), can deal with input/state/output constraints,
and can manage systematically the trade-off between performance and
command activity. \ifthenelse{\acc=1}{}{ Successful applications
of NMPC can be found in many areas, such as automotive engineering,
aerospace engineering, chemical process management, robotics, biomedicine,
etc. }

NMPC requires to solve in real-time an optimization problem, which
is usually nonlinear and may involve a large number of decision variables.
In many cases, this problem is characterized by a high computational
complexity, making its real-time solution difficult or even unfeasible
using standard computers, especially if a large number of variables
is involved. In the last decades, it has been shown that quantum computers,
exploiting quantum mechanical phenomena like superposition and tunneling,
have the potential to significantly reduce the computational complexity
of certain classes of non-convex optimization problems, allowing their
solution with a substantial speedup with respect to classical computers,
providing at the same time better solutions.

This paper presents a quantum optimization approach for NMPC that
may offer relevant improvements in computational time compared to
classical algorithms running on classical computers. The first contribution
of the paper is a novel polynomial formulation of NMPC. One one hand,
this formulation allows a reduction of the computational complexity
required to solve a NMPC optimization problem also on a classical
computer. On the other hand, it serves as the starting point for the
development of a quantum NMPC (QMPC) algorithm, which constitutes
the main contribution of the paper. In particular, the polynomial
formulation is transformed into a Quadratic Unconstrained Binary Optimization
(QUBO) problem. Provided that an efficient quantum computer is available,
the QUBO problem can be solved with a significant computational speedup
(at least quadratic) compared to classical computers, possibly improving
also the solution quality. The proposed QMPC approach has the potential
to become a groundbreaking tool in a near future, when/if effective
and reliable quantum computers will be available, possibly not requiring
too large space occupation, allowing their embedding in dynamic systems
of different sizes.

\ifthenelse{\acc=0}{}{

Although quantum computing is a rapidly evolving field, quantum hardware
is still at an experimental stage, and several issues have to be fully
addressed, like qubit stability, error correction and coherence times.
In this ``dynamic'' context, it is worth mentioning a particular class
of quantum computers, called quantum annealers, whose state of progress
is relatively advanced, allowing the solution of complex optimization
problems of interest in real applications \cite{albash2018demonstration,Boixo2015,kadowaki1998quantum,Morita2008}.
A summary on the state of progress of quantum computer hardware can
be \textcolor{red}{found in ??}.

Two examples are presented in this paper, showing how the proposed
QMPC approach could be an effective tool for problems of practical
interest. The first one is concerned with trajectory planning and
control for autonomous vehicles. The second one regards control of
a paper machine. In both examples, the QMPC approach is compared with
classical methods. In the second example, the runtime of the QMPC/NMPC
optimization problem is evaluated for increasing number $n_{c}$ of
decision variables. This analysis shows that the runtime of the QMPC
optimization problem does not increase with $n_{c}$. }

To the best of our knowledge, the QMPC approach presented in this
paper is the first pure quantum Model Predictive Control (MPC) method
for nonlinear systems. In the literature, we found only another quantum
algorithm for MPC but it is restricted to Linear Time Invariant systems
\cite{Inoue2020}. Also the work \cite{DeGrossi23} deserves to be
mentioned, presenting a quantum-inspired NMPC technique based on a
meta-heuristic approach called Diffusion Monte Carlo. It must be noted
however that quantum-inspired algorithms are significantly different
from pure quantum algorithms: they take inspiration from quantum mechanics
but run on classical hardware and are not specifically designed to
exploit the parallelism of quantum computers.

\ifthenelse{\acc=1}{}{

\subsection{State of progress of quantum computer hardware}

\label{subsec:progr_qh}

Quantum computing is a rapidly evolving field, having the potential
to revolutionize various domains of science, technology, and industry.
The basic unit of quantum computers is the \emph{qubit} (quantum bit),
a two-state system, obeying the laws of quantum mechanics \cite{humble2019quantum}.
Examples include the spin of the electron in which the two levels
can be taken as spin up and spin down, the polarization of a single
photon in which the two states can be taken to be the vertical polarization
and the horizontal polarization \cite{slussarenko2019photonic}. In
a classical system, a bit can be in one state or the other. Instead,
a qubit can be in a superposition of both states simultaneously. This
property, peculiar of quantum systems, allows the so-called quantum
parallelism, which allows the virtual simultaneous exploration of
the entire solution space, unlike classical computers that explore
only one possibility at a time. Such a parallelism is one of the key
features of quantum computers.

We can distinguish between two main kinds of quantum computers: (i)
General-purpose quantum computers: systems that have the capability
to perform a wide range of computations and solve diverse problems
developing a proper quantum algorithm, see, e.g., \cite{Yanofsky_Mannucci_2008,nielsen_chuang_2010,aaronson_2013,preskill_2018}.
(ii) Quantum annealers: specific type of quantum device, designed
to solve a relatively general class of optimization problems, see,
e.g., \cite{albash2018demonstration,Boixo2015,hauke2020perspectives,kadowaki1998quantum,Morita2008}.
The system begins in a quantum superposition of states, representing
potential solutions to the problem. Initially, the system is in the
ground state of the initial Hamiltonian, which promotes quantum fluctuations.
As the system evolves and the problem Hamiltonian is gradually applied
(through an ideally adiabatic evolution), the system should remains
in the ground state, transitioning to the ground state of the problem
Hamiltonian, which corresponds to the optimal solution of the problem.
Both general-purpose computers and quantum annealers have the potential
to outperform classical computers thanks to the aforementioned quantum
parallelism.

Leading companies in the field of general-purpose quantum computers
are IBM, Google, Intel, Microsoft, Honeywell, Xanadu, Rigetti Computing,
Quantinuum, D-Wave, QuEra and IonQ. They have built quantum computers
with tents or even hundreds of qubits, allowing the solution of simple
optimization problems. A leading company in the field of quantum annealers
is D-Wave Systems, which has developed quantum annealers with a growing
number of qubits (over 5000), enabling the solution of increasingly
complex optimization problems. Several companies, including IBM, D-Wave,
Rigetti and IonQ, offer cloud access to their quantum computers, and
the access may be free in the case of non-commercial and short-time
utilization. Some of these companies and other ones, like Google and
Microsoft, offer cloud access to simulators of quantum computers,
which cannot have the same performance of quantum computers but allow
the test of quantum algorithms and a preliminary analysis of the performance
that could be achieved. Note that the availability and features of
these services may evolve and the user must check the providers official
websites for the most recent information on their quantum computing
offers.

As discussed above, the availability of efficient quantum computers
would allow the solution of difficult optimization problems in a significantly
shorter time and with an improved quality with respect to classical
computers. However, quantum hardware is still in the experimental
stage, and issues such as qubit stability, error rates and coherence
times need to be overcome. Researchers and engineers are actively
working to address these challenges and improve the performance and
scalability of quantum computers. Big companies like IBM, Intel, Google,
Microsoft, Honeywell, Xanadu, Rigetti Computing, IonQ and D-Wave are
currently investing significant human and financial resources for
the development of reliable quantum computer hardware. }

\section{Notation }

\label{sec:notation}

A column vector $x\in\mathbb{R}^{n_{x}}$ is denoted by $x=\left(x_{1},\ldots,x_{n_{x}}\right)$.
A row vector $x\in\mathbb{R}^{1\times n_{x}}$ is $x=\left[x_{1},\ldots,x_{n_{x}}\right]=\left(x_{1},\ldots,x_{n_{x}}\right)^{\top}$,
where $\top$ indicates the transpose. The $\ell_{p}$ norm of a vector
$x=\left(x_{1},\ldots,x_{n_{x}}\right)$ is defined as $\left\Vert x\right\Vert _{p}\doteq(\sum_{i=1}^{n_{x}}\left|x_{i}\right|^{p})^{1/p}$,
$p<\infty$, $\left\Vert x\right\Vert _{\infty}\doteq\max_{i}\left|x_{i}\right|$.

Consider a function $g:\mathbb{R}^{n_{c}}\rightarrow\mathbb{R}^{n_{g}}$
of class $\mathbf{C}^{\beta}$. The notation $\mathcal{M}^{(\alpha)}\{g,c\}$
indicates the Maclaurin expansion of $g(c)$ with respect to $c$
truncated at a degree $\alpha\leq\beta$. This expansion is a polynomial
of degree $\alpha$ in $c$ and hence it can be written as
\[
\mathcal{M}^{(\alpha)}\{g,c\}=\mathcal{M}_{\mathcal{C}}^{(\alpha)}\{g\}\,\mu^{(\alpha)}(c)
\]
where $\mu^{(\alpha)}(c)\in\mathbb{R}^{n_{\mu}}$ is a vector containing
all monomials in $c$ up to degree $\alpha$ (with a chosen monomial
order), and $\mathcal{M}_{\mathcal{C}}^{(\alpha)}\{g\}$ is the matrix
containing the coefficients of the Maclaurin expansion. 

Given a vector $x=\left(x_{1},\ldots,x_{n_{x}}\right)$ and two integers
$a$ and $b$ such that $ab=n_{x}$, the reshape operator $RS$ is
defined as 
\[
RS(x,a,b)=\left[\begin{array}{cccc}
x_{1} & x_{a+1} & \cdots & x_{a(b-1)+1}\\
\vdots & \vdots & \ddots & \vdots\\
x_{a} & x_{2a} & \cdots & x_{n_{x}}
\end{array}\right]\in\mathbb{R}^{a\times b}.
\]

The $\mathrm{diag}$ operator is defined as follows. For a square
matrix $X\in\mathbb{R}^{n_{x}\times n_{x}}$, $\mathrm{diag}(X)$
is the column vector of dimension $n_{x}$ containing the diagonal
elements of $X$. For a vector $x\in\mathbb{R}^{n_{x}}$, $\mathrm{diag}(x)$
is the diagonal matrix of dimensions $n_{x}\times n_{x}$ with the
elements of $x$ on the main diagonal.  

\section{Nonlinear MPC }

\label{sec:nmpc}

\subsection{Plant to control}

Consider a continuous-time Multiple-Input-Multiple-Output (MIMO) nonlinear
system (the plant), described by the following state equation:
\begin{equation}
\dot{x}=f_{c}(x,u)\label{eq:plant}
\end{equation}
where $x\in\mathbb{R}^{n_{x}}$ is the system state, $u\in U_{c}\subset\mathbb{R}^{n_{u}}$
is the command input, $U_{c}$ is a bounded set. The state is measured
in real-time with a sampling time $T_{s}$. The resulting measurements
are $x(t_{i}),\;t_{i}=T_{s}i,\;i=0,1,\ldots$ .

The goal is to develop efficient NMPC algorithms finalized at controlling
the system (\ref{eq:plant}). Since the focus of this paper is on
the computational aspects of these algorithms, we do not consider
disturbances and/or model uncertainties. For the same reason, issues
like closed-loop stability, recursive feasibility or state estimation
(if the state is not fully measured) are not investigated. 

\subsection{NMPC formulation}

The NMPC formulation considered in this paper is as follows. At each
time $t_{i}$, the system state $x(t_{i})$ is measured. During the
time interval $[t_{i},t_{i+1})$, an optimal command is computed,
which is applied to the system at time $t_{i+1}$ and kept constant
for $t\in[t_{i+1},t_{i+2})$. The optimal command is obtained by means
of two key operations: prediction and optimization. 

\emph{Prediction.} Let 
\begin{equation}
\hat{x}_{k+1}=f(\hat{x}_{k},\mathfrak{u}_{k})\label{eq:model0}
\end{equation}
be a discrete-time model of the plant (\ref{eq:plant}), where $\hat{x}_{k}\in\mathbb{R}^{n_{x}}$
is the model state, $\mathfrak{u}_{k}\in U_{c}\subset\mathbb{R}^{n_{u}}$
is the input and $f\in\mathbf{C}^{\beta}$, $\beta\geq1$. The model
is obtained using a discretization time $T_{d}$ which in general
may be different from $T_{s}$. At each time $t_{i}$, the state is
predicted $T+1$ steps in the future by iterating $T$ times the model
equation (\ref{eq:model0}), starting from the first prediction $\hat{x}_{1}=f(x(t_{i}),u(t_{i}))$.
The integer $T>1$ is called the \emph{prediction horizon}. For any
$k\in[1,T]$, the predicted state $\hat{x}_{k+1}$ is a function of
the future input sequence $\boldsymbol{\mathfrak{u}}_{1:k}\doteq(\mathfrak{u}_{1},\ldots,\mathfrak{u}_{k})$. 

\emph{Optimization.} The basic idea of NMPC is to compute a command
sequence $\boldsymbol{\mathfrak{u}}_{1:T}^{*}$ such that the predicted
state sequence has an ``optimal behavior'' over the time interval
$[2,T+1]$. The concept of ``optimal behavior'' is captured by the
objective function {\small
\begin{equation}
J\left(\boldsymbol{\mathfrak{u}}\right)\doteq\sum_{k=1}^{T}\mathfrak{u}_{k}^{\top}R\mathfrak{u}_{k}+\sum_{k=2}^{T}\tilde{x}_{k}^{\top}Q\tilde{x}_{k}+\tilde{x}_{T+1}^{\top}\,P\tilde{x}_{T+1}\label{eq:obj}
\end{equation}
}where $\tilde{x}_{k}\doteq\mathfrak{r}_{k}-\hat{x}_{k}$ is the predicted
tracking error, $\mathfrak{r}_{k}\doteq r_{t+k}\in\mathbb{R}^{n_{x}}$
is the reference sequence to track, $\boldsymbol{\mathfrak{u}}\equiv\boldsymbol{\mathfrak{u}}_{1:T}\doteq(\mathfrak{u}_{1},\ldots,\mathfrak{u}_{T})\in\mathbb{R}^{Tn_{u}\times1}$
is the applied command sequence, $T$ is the prediction horizon and
$R,Q,P$ are diagonal weight matrices. The optimal input sequence
is chosen as one minimizing the objective function $J\left(\cdot\right)$,
ensuring that the input and state sequences are consistent with the
model equation and satisfy possible constraints. That is, the optimal
input sequence is computed at each time step $t_{i}\in\mathbb{N}_{0}$
by solving the following optimization problem:
\begin{eqnarray}
 &  & \boldsymbol{\mathfrak{u}}^{*}=\arg\underset{\boldsymbol{\mathfrak{u}}}{\min}\ J\left(\mathfrak{u}\right)\label{eq:opt}\\
 &  & \begin{array}{l}
\textrm{subject to:}\vs{1.5mm}\\
\qquad\hat{x}_{1}=f(x(t_{i}),u(t_{i}))\\
\qquad\hat{x}_{k+1}=f(\hat{x}_{k},\mathfrak{u}_{k}),\;k=1:T\vs{1.5mm}\\
\qquad\hat{x}_{k}\in X_{c},\;k=2:T+1\vs{1.5mm}\\
\qquad\mathfrak{u}_{k}\in U_{c},\;k=1:T.
\end{array}\label{eq:cons}
\end{eqnarray}
The fist two constraints in this problem guarantee that the predicted
state is consistent with the model equation (\ref{eq:model0}). The
set $X_{c}$ accounts for other constraints that may hold on the system
state. The set $U_{c}$ accounts for input constraints. 

\subsection{Input sequence dimension reduction}

The optimization problem (\ref{eq:opt}) is in general non-convex.
Moreover, the command sequence $\boldsymbol{\mathfrak{u}}\doteq(\mathfrak{u}_{1},\ldots,\mathfrak{u}_{T})$
may contain a relatively large number of decision variables. An efficient
method to reduce the number of variables is the Move Blocking technique
\cite{cagiernad2007}: the command sequence $\boldsymbol{\mathfrak{u}}$
is parametrized as
\begin{equation}
\boldsymbol{\mathfrak{u}}=\Gamma c\label{eq:dimred}
\end{equation}
where $\Gamma\in\mathbb{R}^{Tn_{u}\times n_{c}}$ is a full-rank matrix
with $n_{c}<Tn_{u}$, and $c\in\mathbb{R}^{n_{c}}$ is a new command
input sequence with reduced dimension. A sample $\mathfrak{u}_{k}$
of $\boldsymbol{\mathfrak{u}}$ can be obtained using a selection
matrix $S_{k}^{u}$ such that $\mathfrak{u}_{k}=S_{k}^{u}\boldsymbol{\mathfrak{u}}=S_{k}^{u}\Gamma c$. 

\ifthenelse{\acc=1}{}{\begin{example}Suppose that\textcolor{red}{{}
}$\boldsymbol{\mathfrak{u}}\doteq(\mathfrak{u}_{1},\mathfrak{u}_{2},\mathfrak{u}_{3},\mathfrak{u}_{4},\mathfrak{u}_{5})\in\mathbb{R}^{5\times1}$,
$\mathfrak{u}_{k}\in\mathbb{R},\,\forall k$. The parametrization{\small
\[
\boldsymbol{\mathfrak{u}}=\Gamma c=\left[\begin{array}{ccc}
1 & 0 & 0\\
0 & 1 & 0\\
0 & 0 & 1\\
0 & 0 & 1\\
0 & 0 & 1
\end{array}\right]\left[\begin{array}{c}
c_{1}\\
c_{2}\\
c_{3}
\end{array}\right]=\left[\begin{array}{c}
c_{1}\\
c_{2}\\
c_{3}\\
c_{3}\\
c_{3}
\end{array}\right]
\]
}allows a reduction of the number of decision variables from $5$
to $3$. The input is parametrized as a piecewise constant sequence,
with three possible values in the prediction interval. The vector
that selects a sample $\mathfrak{u}_{k}$ is $S_{k}^{u}=[0,\ldots,1,\ldots,0]\in\mathbb{R}^{1\times5}$,
where the unique non-zero element is the one with index $k$. $\square$

\end{example}}

Under the input parametrization (\ref{eq:dimred}), the optimization
problem becomes
\begin{eqnarray}
 &  & c^{*}=\arg\underset{c}{\min}\ J\left(\Gamma c\right)\label{eq:opt-1}\\
 &  & \begin{array}{l}
\textrm{subject to:}\vs{1.5mm}\\
\qquad\hat{x}_{1}=f(x(t_{i}),u(t_{i}))\\
\qquad\hat{x}_{k+1}=f(\hat{x}_{k},S_{k}^{u}\Gamma c),\;k=1:T\vs{1.5mm}\\
\qquad\hat{x}_{k}\in X_{c},\;k=2:T+1\vs{1.5mm}\\
\qquad S_{k}^{u}\Gamma c\in U_{c},\;k=1:T.
\end{array}\label{eq:cons-1}
\end{eqnarray}

\subsection{Receding horizon strategy}

The NMPC feedback command is computed by solving the optimization
problem (\ref{eq:opt-1}) at each time $t_{i}$, according to a so-called
\emph{Receding Horizon Strategy} (RHS). 

\textbf{Receding Horizon Strategy }
\begin{enumerate}
\item In the time interval $[t_{i},t_{i+1})$:
\begin{enumerate}
\item at time $t_{i}$, measure $x(t_{i})$;
\item compute $c^{*}$ by solving (\ref{eq:opt-1}).
\end{enumerate}
\item In the time interval $[t_{i+1},t_{i+2})$:
\begin{enumerate}
\item apply to the plant the first command sample: $u(t)=S_{1}^{u}\Gamma c^{*}$,
$\forall t\in[t_{i+1},t_{i+2})$.
\end{enumerate}
\item Repeat steps 1 and 2 for $t_{i+1},t_{i+2},\ldots$ $\quad\square$
\end{enumerate}
The receding horizon strategy is fundamental to have a feedback control
action, allowing the NMPC algorithm to stabilize unstable systems,
attenuate external disturbances and properly react if sudden changes
occur in the scenario where the plant is operating. Note that this
NMPC/RHS formulation considers a realistic scenario, where the state
is measured at time $t_{i}$, the time interval $[t_{i},t_{i+1})$
is used to compute the optimal command, which is applied during the
next time interval $[t_{i+1},t_{i+2})$.

\section{Polynomial MPC}

Solving the optimization problems (\ref{eq:opt}) and (\ref{eq:opt-1})
analytically is in general hard, and numerical solvers are used. These
kinds of solvers are typically based on the evaluation of the objective
function at a suitable number of points. To obtain the required multi-step
prediction, each objective function evaluation may need to iterate/compute
the state equation $T$ times. In this section, we develop a model
approximation approach based on polynomial expansions, allowing us
to: (i) reduce the time taken to compute the multi-step prediction;
(ii) write the NMPC optimization problem in a form suitable to be
used by quantum annealers, see Section \ref{sec:QMPC}. The approach
of this section is named Polynomial MPC (PMPC).

\subsection{Polynomial prediction algorithm}

Let $\mathcal{M}^{(\alpha)}\{f(\hat{x}_{1},S_{1}\Gamma c),c\}$ be
the Maclaurin expansion of $f(\hat{x}_{1},S_{1}\Gamma c)$ with respect
to $c$, truncated at a polynomial degree $\alpha\leq\beta$.  We
recall that a Maclaurin expansion is a Taylor expansion of a function
around $0$. The generalization of the approach presented in this
paper to the case of Taylor expansion is straightforward. 

The expansion is a polynomial of degree $\alpha$ in $c$ and hence
it can be written as
\[
\mathcal{M}^{(\alpha)}\{f(\hat{x}_{1},S_{1}\Gamma c),c\}=F_{1}^{(\alpha)}(\hat{x}_{1})\mu^{(\alpha)}(c)
\]
where $\mu^{(\alpha)}(c)\in\mathbb{R}^{n_{\mu}}$ is a vector whose
elements are monomials in $c$ up to degree $\alpha$, and $F_{1}^{(\alpha)}(\hat{x}_{1})\doteq\mathcal{M}_{\mathcal{C}}^{(\alpha)}\{f(\hat{x}_{1},S_{1}\Gamma c)\}\in\mathbb{R}^{n_{x}\times n_{\mu}}$
is a matrix containing the coefficients of the expansion. In this
polynomial approach, the prediction at step $k=1$ is computed as
\[
\hat{x}_{2}=F_{1}^{(\alpha)}(\hat{x}_{1})\mu^{(\alpha)}(c).
\]
At step $k=2$, we have
\[
f(\hat{x}_{2},S_{2}\Gamma c)=f(F_{1}^{(\alpha)}\mu^{(\alpha)}(c),S_{2}\Gamma c).
\]
The Maclaurin expansion $\mathcal{M}^{(\alpha)}\{f(F_{1}^{(\alpha)}\mu^{(\alpha)}(c),S_{2}\Gamma c),c\}$
is a polynomial of degree $\alpha$ in $c$, and can thus be written
as
\[
\mathcal{M}^{(\alpha)}\{f(F_{1}^{(\alpha)}\mu^{(\alpha)}(c),S_{2}\Gamma c),c\}=F_{2}^{(\alpha)}(F_{1}^{(\alpha)})\mu^{(\alpha)}(c)
\]
where $F_{2}^{(\alpha)}(F_{1}^{(\alpha)})\doteq\mathcal{M}_{\mathcal{C}}^{(\alpha)}\{f(F_{1}^{(\alpha)}\mu^{(\alpha)}(c),S_{2}\Gamma c)\}\in\mathbb{R}^{n_{x}\times n_{\mu}}$
is the matrix containing the coefficients of the expansion. Note that
$F_{2}^{(\alpha)}$ depends on the previous coefficient matrix $F_{1}^{(\alpha)}$.
The prediction is thus computed as 
\[
\hat{x}_{3}=F_{2}^{(\alpha)}(F_{1}^{(\alpha)})\mu^{(\alpha)}(c).
\]

At a generic time step $k\geq2$, the prediction is obtained as 
\begin{equation}
\hat{x}_{k+1}=F_{k}^{(\alpha)}(F_{k-1}^{(\alpha)})\mu^{(\alpha)}(c)\label{eq:modelp}
\end{equation}
where $F_{k}^{(\alpha)}(F_{k-1}^{(\alpha)})\doteq\mathcal{M}_{\mathcal{C}}^{(\alpha)}\{f(F_{k-1}^{(\alpha)}\mu^{(\alpha)}(c),S_{k}\Gamma c)\}\in\mathbb{R}^{n_{x}\times n_{\mu}}$.
Note that the matrices $F_{k}^{(\alpha)}$, $k=1,\ldots,T$ can be
computed analytically from the model function $f$ by means of two
standard operations: Maclaurin expansion and factorization.

Based on these argumentations, we propose the following prediction
algorithm.

\begin{algorithm} \label{algo:pred}State sequence prediction.

Inputs: current state and command $(x_{t},u_{t})$; future command
sequence $c$.

Output: predicted state sequence $\hat{\mathbf{x}}^{+}\doteq(\hat{x}_{2},\ldots,\hat{x}_{T+1})$.
\begin{enumerate}
\item Coefficient matrix recursive computation:
\begin{enumerate}
\item For $k=1$, compute $\hat{x}_{1}=f(x_{t},u_{t})$ and $F_{1}^{(\alpha)}(\hat{x}_{1})$
$=\mathcal{M}_{\mathcal{C}}^{(\alpha)}\{f(\hat{x}_{1},S_{1}\Gamma c)\}$.
\item For $k=2,\ldots,T$, compute $F_{k}^{(\alpha)}(F_{k-1}^{(\alpha)})$
$=\mathcal{M}_{\mathcal{C}}^{(\alpha)}\{f(F_{k-1}^{(\alpha)}\mu^{(\alpha)}(c),S_{2}\Gamma c)\}$.
\item Define the matrix
\end{enumerate}
\begin{equation}
\begin{array}{l}
\Omega^{(\alpha)}\doteq\left[\begin{array}{c}
F_{1}^{(\alpha)}(\hat{x}_{1})\\
F_{2}^{(\alpha)}(F_{1}^{(\alpha)})\\
\vdots\\
F_{T}^{(\alpha)}(F_{T-1}^{(\alpha)})
\end{array}\right]\in\mathbb{R}^{Tn_{x}\times n_{\mu}}.\end{array}\label{eq:Om}
\end{equation}

\item Prediction. Compute the predicted state sequence $\hat{\mathbf{x}}^{+}\doteq(\hat{x}_{2},\ldots,\hat{x}_{T+1})\in\mathbb{R}^{Tn_{x}\times1}$
as
\begin{equation}
\hat{\mathbf{x}}^{+}=\Omega^{(\alpha)}\mu^{(\alpha)}(c).\quad\square\label{eq:modelm}
\end{equation}
\end{enumerate}
\end{algorithm}

\begin{remark}~The prediction model (\ref{eq:modelp}) and the corresponding
model in matrix form (\ref{eq:modelm}) are factorized: The first
factor contains quantities $F_{k}^{(\alpha)}$ which only depend on
$\hat{x}_{1}$. The second factor is the vector $\mu^{(\alpha)}(c)$
containing the monomials in $c$. The advantages of this factorization
are mainly two: (i) Numerical optimizers that are based on evaluating
the objective function at many points need to compute $\Omega^{(\alpha)}$
only once. The prediction (and subsequently the objective function)
is then evaluated at different points just multiplying this matrix
by $\mu^{(\alpha)}(c)$. For not-too-large polynomial degrees, this
yields a speed improvement in solving the NMPC optimization problem\ifthenelse{\acc=1}{, see Sections \ref{sec:aut_ve} and \ref{sec:pap_mach}}{}.
(ii) As shown in Section \ref{sec:QMPC}, the model (\ref{eq:modelm})
allows us to write the NMPC optimization problem in a form suitable
for quantum annealers, which are a particular class of quantum computers
that are already commercially available \cite{kadowaki1998quantum,Johnson2011,dwave_site}.
$\square$

\end{remark}

\begin{remark}~\label{rmk:standard_lin}In the polynomial approach
developed here, the expansion is performed only with respect (w.r.t.)
to the input $u_{t}$, and not w.r.t. all the arguments of the function
$f(x_{t},u_{t})$. On the other hand, in standard NMPC approaches
based on linearization, the expansion (of degree 1) is typically made
w.r.t. both the state $x_{t}$ and the input $u_{t}$ of the function
$f(x_{t},u_{t})$, implying a lower model accuracy compared to our
approach. $\square$

\end{remark}

\subsection{Polynomial MPC}

Based on the prediction Algorithm \ref{algo:pred}, the Polynomial
MPC (PMPC) approach is now presented. 

For simplicity of notation, let us omit the polynomial degree $\alpha$,
and set $\Omega\doteq\Omega^{(\alpha)}$ and $\mu(c)\doteq\mu^{(\alpha)}(c)$.
We start writing the first term in the right-hand side of (\ref{eq:obj})
as 
\[
\sum_{k=1}^{T}\mathfrak{u}_{k}^{\top}R\mathfrak{u}_{k}=\boldsymbol{\mathfrak{u}}^{\top}\mathbf{W}_{u}\boldsymbol{\mathfrak{u}}=c^{\top}\Gamma^{\top}\mathbf{W}_{u}\Gamma c
\]
where $\mathbf{W}_{u}\doteq\mathrm{diag}(R,\ldots,R)$. Using the
prediction model (\ref{eq:modelm}), the other two terms are given
by
\[
\begin{alignedat}{1} & \sum_{k=2}^{T}\tilde{x}_{k}^{\top}Q\tilde{x}_{k}+\tilde{x}_{T+1}^{\top}P\tilde{x}_{T+1}=\\
 & =(\mathfrak{r}-\Omega\mu(c))^{\top}\mathbf{W}_{x}(\mathfrak{r}-\Omega\mu(c))\\
 & =\mu(c)^{\top}\Omega^{\top}\mathbf{W}_{x}\Omega\mu(c)-2\mathfrak{r}^{\top}\mathbf{W}_{x}\Omega\mu(c)+\mathfrak{r}^{\top}\mathbf{W}_{x}\mathfrak{r}
\end{alignedat}
\]
where $\mathbf{W}_{x}\doteq\mathrm{diag}(Q,\ldots,Q,P)$ and $\boldsymbol{\mathfrak{r}}\doteq(\mathfrak{r}_{2},\ldots,\mathfrak{r}_{T+1})$
is the reference sequence. Hence, we define the objective function
\[
\begin{array}{c}
J_{P}(c)\doteq c^{\top}\Gamma^{\top}\mathbf{W}_{u}\Gamma c+\mu(c)^{\top}\Omega^{\top}\mathbf{W}_{x}\Omega\mu(c)\\
-2\mathfrak{r}^{\top}\mathbf{W}_{x}\Omega\mu(c).
\end{array}
\]
Letting $\mathbf{W}_{\mu}\doteq(S^{c})^{\top}\Gamma^{\top}\mathbf{W}_{u}\Gamma S^{c}+\Omega^{\top}\mathbf{W}_{x}\Omega$,
where $S^{c}$ is the row-selection matrix such that $c=S^{c}\mu(c)$,
the objective function becomes 
\begin{equation}
J_{P}(c)\doteq\mu(c)^{\top}\mathbf{W}_{\mu}\mu(c)-2\mathfrak{r}^{\top}\mathbf{W}_{x}\Omega\mu(c).\label{eq:Jp}
\end{equation}

The PMPC optimal command sequence is computed by solving, at each
time $t_{i}\in\mathbb{N}_{0}$, the optimization problem
\begin{eqnarray}
 &  & c^{*}=\arg\underset{c}{\min}\ J_{P}\left(c\right)\label{eq:opt-2}\\
 &  & \begin{array}{l}
\textrm{subject to:}\vs{1.5mm}\\
\qquad S_{k}^{x}\Omega\mu(c)\in X_{c},\;k=2:T+1\vs{1.5mm}\\
\qquad S_{k}^{u}\Gamma c\in U_{c},\;k=1:T.
\end{array}\label{eq:cons-2}
\end{eqnarray}
where $S_{k}^{x}$ and $S_{k}^{u}$ are row-selection matrices such
that $\hat{x}_{k}=S_{k}^{x}\hat{\mathbf{x}}^{+}=S_{k}^{x}\Omega\mu(c)$
and $\mathfrak{u}_{k}=S_{k}^{u}\boldsymbol{\mathfrak{u}}=S_{k}^{u}\Gamma c$.

The PMPC feedback command is then obtained by applying the Receding
Horizon Strategy described in Section \ref{sec:nmpc}.

\begin{remark}Problem (\ref{eq:opt-2}) is convex if the polynomial
prediction model is input-affine ($\alpha=1$), and $U_{c},U_{x}$
are convex. Note that, in this situation, the PMPC approach is in
general more accurate than standard NMPC approaches based on linearization,
see Remark \ref{rmk:standard_lin}. $\square$

\end{remark}

\section{Quantum MPC with input-affine model and saturation constraints}

\label{sec:QMPC}

In this section, a QUBO (Quadratic Unconstrained Binary Optimization)
formulation of the NMPC optimization problem is developed. This formulation
is of interest since it can be directly used on quantum annealers,
a particular class of quantum computers that are nowadays commercially
available \cite{kadowaki1998quantum,Johnson2011,dwave_site}. Quantum
annealers can solve complex non-convex optimization problems, with
significantly improved performance in terms of computational speed
and/or quality of the solution, compared to classical computers \cite{Boixo2015,Zanca2016,dwave_site}.
A brief summary about quantum computers and, in particular, quantum
annealers can be found in \ifthenelse{\acc=1}{\textcolor{red}{[??]}}{Section \ref{sec:q-ann}}

In the QUBO formulation, the objective function is of the form
\begin{equation}
H=\xi^{\top}\mathbf{Q}\;\xi\label{eq:qubo1}
\end{equation}
where $\mathbf{Q}$ is a matrix of real numbers (to be not confused
with the matrix $Q$ in (\ref{eq:obj})) and $\xi$ is a vector of
binary decision variables. The general procedure that we propose to
obtain the QUBO formulation from the optimization problem (\ref{eq:opt-2})
is as follows.

\textbf{Procedure P1}
\begin{enumerate}
\item \label{enu:bien}Binary encoding of the decision variables. 
\item \label{enu:poma}Polynomial manipulation.
\item \label{enu:qufo}QUBO formulation. $\square$
\end{enumerate}
Here, the procedure is developed for the case of input-affine model
and input saturation constraints. \ifthenelse{\acc=1}{This setting can be in any case relevant in different real-world applications, see the examples presented in Sections \ref{sec:aut_ve} and \ref{sec:pap_mach}.}{}
The more general case of input-polynomial model and polynomial constraints
is treated in \ifthenelse{\acc=1}{[??]}{Section \ref{sec:pmod}}.
Before presenting the procedure, a general binary encoding technique
is proposed for step 1 of the procedure.

\subsection{Binary encoding technique}

The following binary encoding for the decision vector $c=(c_{1},\ldots,c_{n_{c}})$
is considered:
\begin{equation}
c=\underline{c}+C_{b}\Xi\eta\label{eq:bic}
\end{equation}
where $\underline{c}\doteq(\underline{c}_{1},\ldots,\underline{c}_{n_{c}})$
is an offset vector, $C_{b}$ is a diagonal scaling matrix, $\eta$
is a basis vector, and
\[
\Xi\doteq\left[\begin{array}{cccc}
\xi_{1} & \xi_{2} & \cdots & \xi_{n_{b}}\\
\xi_{n_{b}+1} & \xi_{n_{b}+2} & \cdots & \xi_{2n_{b}}\\
\vdots & \vdots & \ddots & \vdots\\
\xi_{(n_{c}-1)n_{b}+1} & \xi_{(n_{c}-1)n_{b}+2} & \cdots & \xi_{n_{c}n_{b}}
\end{array}\right]
\]
is the matrix containing the new binary decision variables $\xi_{j}\in\{0,1\}$,
$j=1,\ldots,m_{0}$, $m_{0}\doteq n_{c}n_{b}$. Note that $\Xi=RS(\xi,n_{b},n_{c})^{\top}$,
where $\xi=(\xi_{1},\ldots,\xi_{m_{0}})$ and $RS$ is the reshape
operator. Other encoding techniques can be\textcolor{red}{{} }found
in \cite{Xavier2023},\textcolor{red}{{} }together with a discussion
of their advantages and drawbacks.

\subsection{Quantum MPC with input-affine model and saturation constraints}

\label{subsec:Quantum-MPC}

In this subsection, the QUBO formulation of the PMPC problem (\ref{eq:opt-2})
is derived assuming $\alpha=1$, corresponding to a prediction model
(\ref{eq:modelm}) affine in the decision vector $c$. Moreover, only
saturation constraints on the components of $c$ are considered. 

\textbf{Assumptions}
\begin{description}
\item [{A1}] \sis{1.5mm}A prediction model (\ref{eq:modelm}) affine in
$c$ is chosen ($\alpha=1$).
\item [{A2}] The decision vector $c$ is subject only to saturation constraints:
\[
c\in\mathcal{C}\doteq\{c\in\mathbb{R}^{n_{c}}:\underline{c}_{i}\leq c_{i}\leq\overline{c}_{i},\,i=1,\ldots,n_{c}\}
\]
where $|\underline{c}_{i}|,|\overline{c}_{i}|<\infty$. 
\item [{A3}] The command $c\in\mathbb{R}^{n_{c}}$ is encoded according
to (\ref{eq:bic}): $c=\underline{c}+C_{b}\Xi\eta$, where $\eta\doteq(2^{n_{b}-1},\ldots,2,0)$
and $C_{b}\doteq\mathrm{diag}(\overline{c}-\underline{c})/(2^{n_{b}}-1)$,
being $\underline{c}$ and $\overline{c}$ the vectors with components
$\underline{c}_{i}$ and $\overline{c}_{i}$, respectively.
\end{description}
The QUBO formulation is obtained by applying Procedure P1.

\emph{P1.\ref{enu:bien}) Binary encoding of the decision variables.
}Using the binary encoding (\ref{eq:bic}), the polynomial vector
$\mu(c)$ in (\ref{eq:Jp}) becomes $\mu(c)=\mu(\underline{c}+C_{b}\Xi\eta)$,
where $\Xi=RS(\xi,n_{b},n_{c})^{\top}$ and $\xi=(\xi_{1},\ldots,\xi_{m_{0}})$
$\in\{0,1\}^{m_{0}}$.

\emph{P1.\ref{enu:poma}) Polynomial manipulation.} Under Assumption
A1, $\mu(c)$ is an affine function: $\mu(c)=(c,1)$. After the binary
encoding operation, the vector is written as
\[
\begin{alignedat}{1}\mu(c) & =(\underline{c}+C_{b}\Xi\eta,1)=M\xi+\underline{\mu}\end{alignedat}
\]
where $\underline{\mu}\doteq(\underline{c},1)$ and $M\doteq\frac{\partial(C_{b}\Xi\eta,1)}{\partial\xi}$. 

\emph{P1.\ref{enu:qufo}) QUBO formulation.} The QUBO formulation
of the PMPC problem (\ref{eq:opt-2}) is the following:
\begin{equation}
\begin{alignedat}{1} & \xi^{*}=\arg\underset{\xi\in\{0,1\}^{m_{0}}}{\min}H(\xi)\\
 & H(\xi)\doteq\xi^{\top}\mathbf{Q}\;\xi
\end{alignedat}
\label{eq:opt-3a}
\end{equation}
where 
\[
\begin{alignedat}{1}\mathbf{Q} & \doteq M^{\top}\mathbf{W}_{\mu}M+\mathrm{diag}(h)\\
h & \doteq2(\underline{\mu}^{\top}\mathbf{W}_{\mu}-\mathfrak{r}^{\top}\mathbf{W}_{x}\Omega)M.
\end{alignedat}
\]
We recall that $\Omega\equiv\Omega^{(\alpha)}$ is defined in (\ref{eq:Om}),
$\mathfrak{r}$ is the reference sequence, $\mathbf{W}_{x}\doteq\mathrm{diag}(Q,\ldots,Q,P)$,
$\mathbf{W}_{\mu}\doteq(S^{c})^{\top}\Gamma^{\top}\mathbf{W}_{u}\Gamma S^{c}+\Omega^{\top}\mathbf{W}_{x}\Omega$,
$\mathbf{W}_{u}\doteq\mathrm{diag}(R,\ldots,R)$, where $Q$, $P$
and $R$ are the weight matrices in the NMPC objective function (\ref{eq:obj}),
$\Gamma$ is a dimension-reduction matrix (\ref{eq:dimred}) and $S^{c}$
is the selection matrix in (\ref{eq:Jp}).

A theorem is now presented, showing that a solution of the optimization
problems (\ref{eq:opt-3a}) can be made arbitrarily close to a solution
of (\ref{eq:opt-2}) by choosing a sufficiently high number $n_{b}$
of binary variables.

\begin{theorem}\label{thm:popt}Let Assumptions A1-A3 hold true.
Let $\xi^{*}$ be a global solution of (\ref{eq:opt-3a}), $\Xi^{*}=RS(\xi^{*},n_{b},n_{c})^{\top}$
and $\breve{c}^{*}=\underline{c}+C_{b}\Xi^{*}\eta$. Let $c^{*}$
be the global solution of (\ref{eq:opt-2}) closest to $\breve{c}^{*}$.
Then, for a sufficiently large $n_{b}$, 
\[
\left\Vert \breve{c}^{*}-c^{*}\right\Vert _{\infty}\leq\frac{\left\Vert \overline{c}-\underline{c}\right\Vert _{\infty}}{2^{n_{b}}-1}.
\]
Moreover, a finite positive constant $\gamma_{P}$ (independent of
$n_{b}$) exists such that
\[
\left|J_{P}(\breve{c}^{*})-J_{P}(c^{*})\right|\leq\frac{\gamma_{P}\left\Vert \overline{c}-\underline{c}\right\Vert _{\infty}}{2^{n_{b}}-1}.
\]

\end{theorem}

\textbf{Proof. }Under the assumption $\alpha=1$, we have that $\mu(c)=(c,1)=M\xi+\underline{\mu}$,
where $\underline{\mu}\doteq(\underline{c},1)$ and $M\doteq\frac{\partial(C_{b}\Xi\eta,1)}{\partial\xi}$.
Then, the objective function $J_{P}$ in (\ref{eq:Jp}) becomes 
\begin{equation}
\begin{alignedat}{1}\breve{J}_{P}(\xi) & =\mu^{\top}\mathbf{W}_{\mu}\mu-2\mathfrak{r}^{\top}\mathbf{W}_{x}\Omega\mu\\
 & =(M\xi+\underline{\mu})^{\top}\mathbf{W}_{\mu}(M\xi+\underline{\mu})\\
 & -2\mathfrak{r}^{\top}\mathbf{W}_{x}\Omega(M\xi+\underline{\mu})\\
 & =\xi^{\top}M^{\top}\mathbf{W}_{\mu}M\xi\\
 & +2(\underline{\mu}^{\top}\mathbf{W}_{\mu}-\mathfrak{r}^{\top}\mathbf{W}_{x}\Omega)M\xi\\
 & +\underline{\mu}^{\top}\mathbf{W}_{\mu}\underline{\mu}-2\mathfrak{r}^{\top}\mathbf{W}_{x}\Omega\underline{\mu}.
\end{alignedat}
\label{eq:Jp_}
\end{equation}
Note that $\breve{J}_{P}$ corresponds to the objective function $J_{P}$
with a discretized domain. We thus use different symbols for the two
functions. 

Defining $h\doteq2(\underline{\mu}^{\top}\mathbf{W}_{\mu}-\mathfrak{r}^{\top}\mathbf{W}_{x}\Omega)M$,
we have that
\[
\begin{alignedat}{1} & 2(\underline{\mu}^{\top}\mathbf{W}_{\mu}-\mathfrak{r}^{\top}\mathbf{W}_{x}\Omega)M\xi=h\xi\\
 & =\sum_{i=1}^{m_{0}}h_{i}\xi_{i}=\sum_{i=1}^{m_{0}}h_{i}\xi_{i}^{2}=\xi^{\top}\mathrm{diag}(h)\,\xi
\end{alignedat}
\]
where we have used the property that $\xi_{i}\in\{0,1\}$ $\Rightarrow$
$\xi_{i}^{2}=\xi_{i}$. From (\ref{eq:Jp_}) and the definition of
$h$, it follows that
\[
\begin{alignedat}{1}\breve{J}_{P}(\xi) & =\xi^{\top}M^{\top}\mathbf{W}_{\mu}M\xi+\xi^{\top}\mathrm{diag}(h)\,\xi\\
 & +\underline{\mu}^{\top}\mathbf{W}_{\mu}\underline{\mu}-2\mathfrak{r}^{\top}\mathbf{W}_{x}\Omega\underline{\mu}\\
 & =\xi^{\top}(M^{\top}\mathbf{W}_{\mu}M+\mathrm{diag}(h))\,\xi\\
 & +\underline{\mu}^{\top}\mathbf{W}_{\mu}\underline{\mu}-2\mathfrak{r}^{\top}\mathbf{W}_{x}\Omega\underline{\mu}\\
 & =H(\xi)+\underline{\mu}^{\top}\mathbf{W}_{\mu}\underline{\mu}-2\mathfrak{r}^{\top}\mathbf{W}_{x}\Omega\underline{\mu}\\
 & =H(\xi)+\mathrm{const}.
\end{alignedat}
\]

Consider now that $\xi^{*}$ is a global minimum of $H$ and thus
it is also a global minimum of $\breve{J}_{P}$, since the difference
between the two functions is a constant. Let $\Xi^{*}=RS(\xi^{*},n_{b},n_{c})^{\top}$
be the corresponding matrix, $\breve{c}^{*}=\underline{c}+C_{b}\Xi^{*}\eta$,
and let $c^{*}$ be the global minimum of $J_{P}$ closest to $\breve{c}^{*}$. 

We observe that the encoding $c=\underline{c}+C_{b}\Xi\eta$, with
$\eta\doteq(2^{n_{b}-1},\ldots,2,0)$ and $C_{b}\doteq\mathrm{diag}(\overline{c}_{1}-\underline{c}_{1},\ldots,\overline{c}_{n_{c}}-\underline{c}_{n_{c}})/(2^{n_{b}}-1)$,
defines a grid in the domain $\mathcal{C}$ of the decision vector
$c$, and each cell of this grid is a cuboid with side lengths $\frac{\overline{c}_{i}-\underline{c}_{i}}{2^{n_{b}}-1}$.
It can be observed that the side lengths tend to $0$ as $n_{b}\rightarrow\infty$.
Hence, since $J_{P}$ is Lipschitz continuous on $\mathcal{C}$, a
sufficiently large $n_{b}$ exists, such that
\begin{equation}
\left\Vert \breve{c}^{*}-c^{*}\right\Vert _{\infty}\leq\frac{\left\Vert \overline{c}-\underline{c}\right\Vert _{\infty}}{2^{n_{b}}-1}\label{eq:sol_diff}
\end{equation}
where $\underline{c}$ and $\overline{c}$ are the vectors with components
$\underline{c}_{i}$ and $\overline{c}_{i}$, respectively. 

Moreover, from Lipschitz continuity of $J_{P}$, it follows that a
finite constant $\gamma_{P}$ exists such that
\begin{equation}
\begin{array}{c}
\left|J_{P}(\breve{c}^{*})-J_{P}(c^{*})\right|\leq\gamma_{P}\left\Vert \breve{c}^{*}-c^{*}\right\Vert _{\infty}\leq\gamma_{P}\frac{\left\Vert \overline{c}-\underline{c}\right\Vert _{\infty}}{2^{n_{b}}-1}.\end{array}\label{eq:J_diff}
\end{equation}

Since the minimizers of $H$ are the same as the minimizers of $\breve{J}_{P}$,
the claim follows from (\ref{eq:sol_diff}) and (\ref{eq:J_diff}).
$\square$

\section{Quantum MPC with input-polynomial model and polynomial constraints}

\label{sec:pmod}

In this section, we relax the assumption of affine model made in Section
\ref{sec:QMPC}, and treat the case where the prediction model is
polynomial in the decision variables. We start considering saturation
constraints (Subsection \ref{subsec:poly_case}) and then we move
to polynomial constraints (Subsection \ref{subsec:con_case}).

\subsection{Quantum MPC with input-polynomial model and saturation constraints}

\label{subsec:poly_case}

In this subsection, we discuss the case where the prediction model
is polynomial in the decision variables and input-saturation constraints
hold.

\textbf{Assumptions}
\begin{description}
\item [{A1b}] \sis{1.5mm}A polynomial prediction model (\ref{eq:modelm})
with degree $\alpha>1$ is chosen.
\item [{A2b}] Assumptions A2-A3 hold.
\end{description}
The QUBO formulation is obtained by applying Procedure P1.

\emph{P1.\ref{enu:bien}) Binary encoding of the decision variables.}
Replacing $c$ with its binary encoding (\ref{eq:bic}), the polynomial
vector $\mu(c)$ in (\ref{eq:Jp}) becomes $\mu(\underline{c}+C_{b}\Xi\eta)=(\mu_{1},\ldots,\mu_{n_{\mu}})$,
where each component $\mu_{i}$ of $\mu$ is a polynomial of degree
$\alpha$ in $\xi^{\{0\}}\doteq(\xi_{1},\ldots,\xi_{m_{0}})$, $m_{0}\doteq n_{c}n_{b}$.
In particular, each component of $\mu$ is a superposition of monomials
of the form
\begin{equation}
\xi_{1}^{\nu_{1}}\xi_{2}^{\nu_{2}}\ldots\xi_{m_{0}}^{\nu_{m_{0}}}\label{eq:mon1}
\end{equation}
where $\nu_{1},\ldots,\nu_{m_{0}}\in\{0,\ldots,\alpha\}$, $\sum_{j=1}^{m_{0}}\nu_{j}\leq\alpha$. 

\emph{P1.\ref{enu:poma}) Polynomial manipulation.} An affine representation
of $\mu(\underline{c}+C_{b}\Xi\eta)$ is obtained using the following
procedure. \vs{0.5em}

\textbf{Procedure P2}

Input: Polynomial $\mu$ of degree $\alpha$ in $\xi\doteq(\xi_{1},\ldots,\xi_{l_{1}})$,
where $\xi$ is a generic binary vector of length $l_{1}$. 

Outputs: 
\begin{enumerate}
\item Polynomial $\mu_{d}$ of any desired degree $\alpha_{d}<\alpha$ in
the augmented vector $(\xi,\xi^{\mathrm{new}})$, such that $\mu_{d}=\mu$
for a suitable $\xi^{\mathrm{new}}$, where $\xi^{\mathrm{new}}\doteq(\xi_{l_{1}+1},\ldots,\xi_{l_{2}})$
is a vector of new binary variables used to reduce the polynomial
degree.
\item Penalty matrix $\Upsilon\in\mathbb{R}^{l_{2}\times l_{2}}$ associated
to $\xi^{\mathrm{new}}$. 
\end{enumerate}
Procedure:
\begin{enumerate}
\item For any binary variable $\xi_{j}\in\{0,1\}$, we have that $\xi_{j}^{\nu}=\xi_{j}$,
$\forall\nu\geq1$. This property allows us to eliminate from the
polynomial vector $\mu$ all powers of the variables larger than $1$.
Hence, a monomial of the form (\ref{eq:mon1}) can be written as $\xi_{1}^{\nu_{1}}\xi_{2}^{\nu_{2}}\ldots\xi_{l_{1}}^{\nu_{l_{1}}}=\xi_{1}\xi_{2}\ldots\xi_{l-1}$.
\item \label{enu:dr2}For any product $\xi_{i}\xi_{j}$, $i,j\leq l_{1}$
a new binary variable $\xi_{l}$, $l>l_{1}$, is introduced, such
that $\xi_{l}=\xi_{i}\xi_{j}$, allowing us to eliminate from a polynomial
the products between pairs of variables \cite{gabor2022}. The equality
$\xi_{l}=\xi_{i}\xi_{j}$ is enforced by adding to the objective function
the penalty $\Lambda_{p}(3\xi_{l}-2\xi_{l}\xi_{i}-2\xi_{l}\xi_{j}+\xi_{i}\xi_{j})$,
where $\Lambda_{p}>0$ is a sufficiently large coefficient. It can
be easily seen that $\xi_{l}=\xi_{i}\xi_{j}$ if and only if the penalty
is null. Note that the penalty is quadratic and can thus be written
in the QUBO form. By repeating this step a sufficient number of times,
the degree of a monomial $\xi_{1}\xi_{2}\ldots\xi_{l_{1}}$ can be
reduced to any desired non-negative value $\alpha_{d}<\alpha$.
\item A penalty matrix $\Upsilon\in\mathbb{R}^{l_{2}\times l_{2}}$ is constructed
as follows: According to step \ref{enu:dr2}, for each product $\xi_{i}\xi_{j}$
replaced by a new variable $\xi_{l}$, assign $\Upsilon_{ij}=1$,
$\Upsilon_{li}=-2$, $\Upsilon_{lj}=-2$, $\Upsilon_{ll}=3$. All
the other elements of $\Upsilon$ are null. $\square$
\end{enumerate}

Steps 1 and 3 of this procedure are straightforward. Step 2 is relatively
simple from a conceptual point of view but computational complexity
issues may arise, possibly leading to long runtimes.\textcolor{red}{{}
}A ``low complexity'' algorithm for executing the procedure is under
development. 

By means of Procedure P2, the polynomial vector $\mu(\xi)\equiv\mu(\underline{c}+C_{b}\Xi\eta)$
is transformed into a vector $\mu_{d}(\xi)\in\mathbb{R}^{n_{\mu}}$
affine in the augmented vector $\xi\doteq(\xi^{\{0\}},\xi^{\{1\}})$,
where $\xi^{\{0\}}\doteq(\xi_{1},\ldots,\xi_{m_{0}})=RS(\Xi,m_{0},1)$
and $\xi^{\{1\}}\doteq(\xi_{m_{0}+1},\ldots,\xi_{m_{1}})$ is a vector
of new binary variables, introduced to eliminate the products $\xi_{i}\xi_{j}$
according to step \ref{enu:dr2} of Procedure P2.  The resulting
affine vector is written as
\begin{equation}
\mu_{d}=M^{\{1\}}\xi+\underline{\mu}^{\{1\}}\label{eq:mu_aff}
\end{equation}
where $\xi\doteq(\xi^{\{0\}},\xi^{\{1\}})$, and $\underline{\mu}^{\{1\}}$
and $M^{\{1\}}$ are a suitable vector and matrix. 

\emph{P1.\ref{enu:qufo}) QUBO formulation.} Using the affine vector
(\ref{eq:mu_aff}), the QUBO formulation of the PMPC problem (\ref{eq:opt-2})
is 
\begin{equation}
\begin{alignedat}{1} & \xi^{*}=\arg\underset{\xi\in\{0,1\}^{m_{1}}}{\min}H(\xi)\\
 & H(\xi)\doteq\xi^{\top}\mathbf{Q}^{\{1\}}\;\xi
\end{alignedat}
\label{eq:opt-3b}
\end{equation}
where 
\[
\begin{alignedat}{1}\mathbf{Q}^{\{1\}} & \doteq M^{\{1\}\top}\mathbf{W}_{\mu}M^{\{1\}}+\mathrm{diag}(h^{\{1\}})+\Lambda^{\{1\}}\Upsilon^{\{1\}}\\
h^{\{1\}} & \doteq2(\underline{\mu}^{\{1\}\top}\mathbf{W}_{\mu}-\mathfrak{r}^{\top}\mathbf{W}_{x}\Omega)M^{\{1\}\top}
\end{alignedat}
\]
and $\Upsilon^{\{1\}}$ is the penalty matrix associated to $\xi^{\{1\}}$
according to Procedure P2, and $\Lambda^{\{1\}}>0$ is a sufficiently
large coefficient. 

\begin{remark}In the worst-case situation, the number $m_{1}$ of
binary decision variables $\xi_{i}$ in (\ref{eq:opt-3b}) is exponential
in the degree $\alpha$ of the polynomial prediction model: $m_{1}=O(e^{\alpha})$.
It is thus convenient to choose a ``low'' degree for this model. %
{} %

\end{remark}

\subsection{Quantum MPC with input-polynomial model and polynomial constraints}

\label{subsec:con_case}

In this subsection, we discuss the general case where the prediction
model is polynomial in the decision variables and polynomial constraints
hold on the system state and input.

\textbf{Assumptions}
\begin{description}
\item [{A1c}] \sis{1.5mm}Assumptions A1b, A2 and A3 hold.
\item [{A2c}] Polynomial input and state constraints must be satisfied.
\end{description}
In QUBO formulations, possible constraints on the decision variables
are imposed by means of penalty functions. We now show how equality
and inequality constraints can be converted into suitable penalty
functions.

\emph{Equality polynomial constraint}. A polynomial equality constraint
on the command $c$ can be written as $\mathcal{P}(c)=0$, where $\mathcal{P}(\cdot)$
is a polynomial. A polynomial equality constraint on the predicted
state sequence $\hat{\mathbf{x}}^{+}\doteq(\hat{x}_{2},\ldots,\hat{x}_{T+1})$
can be written as $\mathcal{P}_{s}(\hat{\mathbf{x}}^{+})=0$, where
$\mathcal{P}_{s}(\cdot)$ is a polynomial. According to (\ref{eq:modelm}),
we have that $\hat{\mathbf{x}}^{+}=\Omega\mu(c)$, where $\mu(c)$
is a vector containing monomials in the command vector $c$ up to
degree $\alpha$. The constraint equation thus becomes $\mathcal{P}_{s}(\Omega\mu(c))=0$.
Since $\mu(c)$ is a polynomial, this equation is polynomial in $c$
and can be written in the form $\mathcal{P}(c)=0$, where $\mathcal{P}(\cdot)$
is a polynomial. 

\emph{Inequality polynomial constraint}. Following a similar reasoning,
a polynomial inequality constraint on the command $c$ or on the state
sequence can be expressed as $\mathcal{P}_{d}(c)\leq0$, where $\mathcal{P}_{d}(\cdot)$
is a polynomial. The inequality is satisfied if and only if $\mathcal{P}_{d}(c)=-s$,
for some slack variable $s\geq0$. Hence, any polynomial inequality
constraint on the command or on the state sequence can be expressed
by an equality of the form $\mathcal{P}(c,s)=0$, where $\mathcal{P}(c,s)\doteq\mathcal{P}_{d}(c)+s$
and $s\geq0$. 

\emph{Generic polynomial constraint}. From the above argumentations,
any polynomial constraint on the command or on the state sequence
can be expressed as $\mathcal{P}(c,s)=0$, where $s=0$ (equality
constraint) or $s\geq0$ (inequality constraint). In order to convert
a constraint into a penalty term, we observe that the equation $\mathcal{P}(c,s)=0$
is satisfied if it admits a solution and $(c,s)$ is a minimizer of
the function $\mathcal{P}(c,s)^{2}$. In order to include the function
$\mathcal{P}(c,s)^{2}$ in the QUBO problem as a penalty term, we
apply Procedure P1.

\emph{P1.\ref{enu:bien}) Binary encoding of the decision variables.
}By assumption A2, $c$ is encoded according to (\ref{eq:bic}). The
variable $s$ can be encoded using the same technique. In particular,
considering that $s\geq0$, we choose $s=C_{s}\xi^{\{2\}\top}\eta$,
where $C_{s}>0$ is a scaling factor and $\xi^{\{2\}}\doteq(\xi_{m_{1}+1},\ldots,\xi_{m_{2}})$
is a vector of new binary variables. The function $\mathcal{P}(\underline{c}+C_{b}\Xi\eta,C_{s}\xi^{\{2\}\top}\eta)^{2}$
is a polynomial in the vector $(\xi^{\{0\}},\xi^{\{2\}})$, where
we recall that $\xi^{\{0\}}\doteq(\xi_{1},\ldots,\xi_{m_{0}})=RS(\Xi,m_{0},1)$.

\emph{P1.\ref{enu:poma}) Polynomial manipulation.} The degree of
the polynomial $\mathcal{P}(\underline{c}+C_{b}\Xi\eta,C_{s}\xi^{\{2\}\top}\eta)^{2}$
is reduced to $2$ by means of Procedure P2. The first output of the
procedure is a quadratic polynomial, which can be written in the QUBO
form as
\begin{equation}
\xi^{\top}\mathbf{Q}_{c}^{\{3\}}\xi\label{eq:Qidef}
\end{equation}
where $\xi\doteq(\xi^{\{0\}},\xi^{\{1\}},\xi^{\{2\}},\xi^{\{3\}})$
and $\xi^{\{3\}}\doteq(\xi_{m_{2}+1},\ldots,\xi_{m_{3}})$ is a vector
of new binary variables, introduced to eliminate the products $\xi_{i}\xi_{j}$
according to step \ref{enu:dr2} of Procedure P2. The second output
of Procedure P2 is a penalty matrix $\Upsilon^{\{3\}}\in\mathbb{R}^{m_{3}\times m_{3}}$
associated to $\xi^{\{3\}}$.

If $\ell$ polynomial constraints have to be satisfied, these operations
are repeated for each of them, giving the matrices $\mathbf{Q}_{c}^{\{3\}},\Upsilon^{\{3\}},\mathbf{Q}_{c}^{\{5\}},\Upsilon^{\{5\}},\ldots,\mathbf{Q}_{c}^{\{\bar{m}\}},\Upsilon^{\{\bar{m}\}}$,
$\bar{m}\doteq m_{2\ell+1}$.

\emph{P1.\ref{enu:qufo}) QUBO formulation.} Suppose that $\ell$
polynomial constraints have to be satisfied. The QUBO formulation
of the PMPC problem (\ref{eq:opt-2}) is the following:
\begin{equation}
\begin{alignedat}{1} & \xi^{*}=\arg\underset{\xi\in\{0,1\}^{\bar{m}}}{\min}H(\xi)\\
 & H(\xi)\doteq\xi^{\top}\mathbf{Q}^{\{\bar{m}\}}\;\xi
\end{alignedat}
\label{eq:opt-3c}
\end{equation}
where
\[
\mathbf{Q}^{\{\bar{m}\}}\doteq\sum_{i\in\{1,3,\ldots,\bar{m}\}}\mathbf{Q}_{a}^{\{i\}}
\]
{\small
\[
\begin{alignedat}{1}\mathbf{Q}_{a}^{\{1\}} & \doteq\left[\begin{array}{cc}
\mathbf{Q}^{\{1\}} & \mathbf{0}\\
\mathbf{0} & \mathbf{0}
\end{array}\right]\in\mathbb{R}^{\bar{m}\times\bar{m}}\\
\mathbf{Q}_{a}^{\{i\}} & \doteq\left[\begin{array}{cc}
\mathbf{Q}_{c}^{\{i\}} & \mathbf{0}\\
\mathbf{0} & \mathbf{0}
\end{array}\right]+\Lambda^{\{i\}}\left[\begin{array}{cc}
\Upsilon^{\{i\}} & \mathbf{0}\\
\mathbf{0} & \mathbf{0}
\end{array}\right]\in\mathbb{R}^{\bar{m}\times\bar{m}},i>1
\end{alignedat}
\]
}$\Upsilon^{\{i\}}$ are the penalty matrices associated to $\xi^{\{i\}}$
according to Procedure P2, $\Lambda^{\{i\}}>0$ are sufficiently large
coefficients, and $\mathbf{0}$ (bold style) indicates a null matrix
of compatible dimensions.

It must be remarked that, in the case of ``large'' polynomial degree
of the model and/or ``large'' number of polynomial constraints, the
number $\bar{m}$ of decision variables may be considerable, compromising
the possibility of solving the QUBO problem on the quantum annealers
that are currently available. Novel approaches are currently being
developed to mitigate this issue.\textcolor{red}{{} }%

\section{Quantum annealers}

\label{sec:q-ann}

Quantum annealers \cite{kadowaki1998quantum,Johnson2011} are special-purpose
quantum computers suitable to solve QUBO problems like \eqref{eq:opt-3a}
(or \eqref{eq:opt-3b}, \eqref{eq:opt-3c}) with a significant computational
speedup (quadratic) compared to classical computers, possibly improving
also the solution quality \cite{Boixo2015,Zanca2016}. In this section,
we summarize what is a quantum annealer and how it works. Defining

\[
\sigma=(\sigma_{1},\ldots,\sigma_{m})\doteq2\xi-I\in\{-1,1\}^{m}\subset\mathbb{Z}^{m\times1},
\]
the objective function of a generic QUBO problem can be written as
\[
\begin{alignedat}{1}H & =\xi^{\top}\mathbf{Q}\;\xi=\frac{1}{4}(\sigma+I)^{\top}\mathbf{Q}\;(\sigma+I)\\
 & =\frac{1}{4}\sigma^{\top}\mathbf{Q}\sigma+\frac{1}{2}I^{\top}\mathbf{Q}\sigma+\frac{1}{4}=H_{F}+\frac{1}{4}
\end{alignedat}
\]

The function $H_{F}$ defined as 
\begin{equation}
H_{F}\doteq\sigma^{\top}J_{F}\sigma+h_{F}\sigma,\;J_{F}\doteq\frac{1}{4}\sigma^{\top}\mathbf{Q},\;h_{F}\doteq\frac{1}{2}I^{\top}\mathbf{Q}\label{eq:hamf}
\end{equation}
is called transverse-field Ising Hamiltonian and represents the energy
of a physical system consisting of $m$ interacting particles, described
by two-state variables $\sigma_{i}$, $i=1,\ldots,m$ (e.g., spin
of electrons, polarization of photons, superconducting loops, etc.).
A Quantum Annealer (QA) is indeed a lattice of $m$ interacting nodes,
called qubits, described by two-state variables, which obey the laws
of quantum mechanics. They are physically implemented by means of
superconducting loops \cite{hauke2020perspectives,Johnson2011}. While
a bit of a classical computer is allowed to be in one of two possible
states, a qubit can be in a superposition of both states simultaneously.
This property, peculiar of quantum systems, enables the so-called
quantum parallelism, where the computer can explore simultaneously
multiple solutions of a problem, unlike classical computers that explore
only one solution at a time. Such a parallelism is one of the key
features of QAs. Another important feature is quantum tunneling, which
allows the computer to bypass energy barriers, thus reducing the issue
of trapping in local solutions. The main steps accomplished by a QA
to solve a QUBO-like problem are as follows.

\subsubsection*{Quantum Annealing Process}
\begin{enumerate}
\item \label{enu:ini}\textit{Initialization:} The QA system is initialized
in a superposition of states, representing all the possible configurations
of the optimization problem. It means that each qubit is in a superposition
of its basis state $0$ and $1$. The corresponding initial Hamiltonian
is denoted by $H_{I}$. %
\item \textit{Adiabatic evolution:} The Hamiltonian is continuously modified
from $H_{I}$, whose ground state is easily configured as initial
state, to the final Hamiltonian $H_{F}$ in \eqref{eq:hamf}, which
encodes the optimization problem. The evolution is controlled in an
adiabatic manner, in such a way that the QA remains in the ground
state during the whole process, also avoiding local minima thanks
to quantum tunneling. The evolution is described by $H_{QA}=\varphi(t)H_{I}+(1-\varphi(t))H_{F}$,
where $H_{QA}$ is the QA Hamiltonian and $\varphi(t)$ is a continuous
function of time such that $\varphi(0)=1$ and $\varphi(T_{A})=0$,
being $T_{A}>0$ called the annealing time. %
\item \textit{Measurement:} Once the evolution process is complete, the
final spin vector $\sigma^{*}$ is measured. This vector corresponds
to a minimum of the Hamiltonian which, at the end of the process,
is $H_{F}=H-1/4$. It follows that the solution of the QUBO problem
\eqref{eq:opt-3a} is $\xi^{*}=(\sigma^{*}+I)/2$. 
\item \label{enu:post}\textit{Post-processing:} The obtained solution might
not be perfect due to noise and defects in the quantum hardware. Post-processing
techniques like classical optimization algorithms can be used to refine
the solution (e.g., simulated annealing or tabu search). 
\item \textit{Repetitions:} The whole procedure (steps \ref{enu:ini} -
\ref{enu:post}) can be repeated several times and the best result
selected. $\Square$ 
\end{enumerate}
After the initialization step, the system is a superposition condition,
guaranteeing a simultaneous and parallel exploration of the solution
space, giving an exponential speedup with respect to a classical computer,
where the solutions must be explored one at a time. However, the evolution
from the initial to the final Hamiltonian cannot be arbitrarily fast.
The Adiabatic Theorem states that a quantum system starting from a
ground state persists in a ground state, provided that the change
in time of the Hamiltonian is sufficiently slow: Let $T_{A}$ be the
annealing time, i.e., the time taken to change the Hamiltonian from
$H_{I}$ to $H_{F}$. According to the Adiabatic Theorem, the ground
state is maintained during the change if

\begin{equation}
T_{A}\geq\frac{1}{\min_{t\in[0,T_{A}]}\Delta E(t)^{2}}\label{eq:adiab}
\end{equation}
where $\Delta E$ is the gap between the two lowest energy levels
of the QA. The combination of the exponential speedup in the initialization
step and the limit imposed by the Adiabatic Theorem yield in any case
a significant computational complexity reduction with respect to classical
computers. Indeed, despite it is notoriously difficult to analyze
the runtime of adiabatic optimization algorithms, several works in
the literature show that at least a quadratic speedup with respect
to classical computers can be guaranteed \cite{Boixo2015,Zanca2016}.

If inequality \eqref{eq:adiab} is not satisfied, there is a non-null
probability for the QA to jump from the lowest energy level to the
second lowest level, or even to higher energy levels, which correspond
to sub-optimal solutions of the optimization problem. This gives rise
to a trade-off between computational speed and quality of the solution:
sub-optimal solutions can be found, which require a shorter annealing
time with respect to an optimal solution but are satisfactory in practice.

A leading company in the field of QAs is D-Wave Systems \cite{dwave_site},
which is developing QAs with a growing number of qubits (more than
5000), enabling the solution of increasingly complex optimization
problems. Quantum annealing is still an evolving technology, and there
are ongoing efforts worldwide to improve its performance and reliability.

\bibliographystyle{ieeetr}
\bibliography{mpc,quantum_comp}

\end{document}